%% file: main.tex
\documentclass{article}
\usepackage{spconf,amsmath,amssymb, graphicx,url,times, pifont}
\usepackage{float}
\usepackage{multicol}
\usepackage{makecell}
\usepackage{subcaption}
\usepackage{amsfonts}
\usepackage{mathtools}
\usepackage{multirow}
\usepackage{bm}
\usepackage{nicefrac}
\usepackage{tabularx}
\usepackage{booktabs}
\usepackage{xcolor}
\usepackage{amsthm}
\usepackage{extarrows}
\usepackage{balance}

\newcommand{\eref}[1]{(\ref{#1})}

\newcommand{\fref}[1]{Fig.~\ref{#1}}
\newcommand{\tref}[1]{Table~\ref{#1}}

\def\bphi{\bm{\phi}}
\def\btheta{\bm{\theta}}
\def\bvarphi{\bm{\varphi}}

\title{Meta-AF Echo Cancellation for Improved Keyword Spotting}
\name{Jonah Casebeer\thanks{$^1$\textit{https://jmcasebeer.github.io/metaaf/kws-af}} \qquad
Junkai Wu \qquad 
Paris Smaragdis 
}%
\address{
University of Illinois at Urbana-Champaign\\
}

\begin{document}

\ninept
\maketitle

\begin{sloppy}

\begin{abstract}
Adaptive filters~(AFs) are vital for enhancing the performance of downstream tasks, such as speech recognition, sound event detection, and keyword spotting.
However, traditional AF design prioritizes isolated signal-level objectives, often overlooking downstream task performance. 
This can lead to suboptimal performance.
Recent research has leveraged meta-learning to automatically learn AF update rules from data, alleviating the need for manual tuning when using simple signal-level objectives.
This paper improves the Meta-AF~\cite{casebeer2022meta} framework by expanding it to support end-to-end training for arbitrary downstream tasks. We focus on classification tasks, where we introduce a novel training methodology that harnesses self-supervision and classifier feedback.
We evaluate our approach on the combined task of acoustic echo cancellation and keyword spotting. Our findings demonstrate consistent performance improvements with both pre-trained and joint-trained keyword spotting models across synthetic and real playback. Notably, these improvements come without requiring additional tuning, increased inference-time complexity, or reliance on oracle signal-level training data.
\end{abstract}


\input{sections/intro}

\input{sections/background}
\input{sections/methods}

\input{sections/experimental_setup}
\input{sections/results}
\input{sections/conclusion}

\bibliographystyle{IEEEbib}
\bibliography{refs23}
\end{sloppy}
\end{document}

%% file: sections/intro.tex
\section{Introduction}
Adaptive filters~(AFs) are essential for many smart systems, including automatic speech recognition, sound event detection, and keyword spotting~\cite{haeb2019speech}. In these applications, AFs serve as key preprocessing steps and are designed to enhance the performance of their downstream components by removing noise, reverberation, and other distortions from their input signals. Traditionally, AFs have been hand-tuned for optimizing signal-level objectives, such as minimizing mean-squared error, using techniques like recursive least squares~\cite{Widrow1985, mathews1991adaptive, haykin2008adaptive}. However, recent advancements in deep learning have introduced methods for controlling pre-built AF update rules~\cite{casebeer2021nice, ivry2022deep, haubner2022end}, or for learning entirely new update rules from scratch ~\cite{casebeer2021auto, casebeer2022meta}. Existing techniques often do not take into account the downstream task and may not lead to the best performance in practice.

Several pioneering works demonstrated that incorporating downstream task knowledge into AF control, through manual unification schemes, could enhance performance~\cite{seltzer2004likelihood, shi2006phase, zhao2007closely}. Such improvements necessitate a complex, task/model-dependent derivation process, as well as feedback from the downstream model to the AF at test time. This line of work inspired many DNN-based approaches. These DNN approaches can be broadly grouped into two categories. The first explicitly decouples preprocessing and downstream operations, treating them as sequential components within a larger DNN pipeline~\cite{heymann2017beamnet, ochiai2017unified, haeb2019speech, zhang2022end}. This is highly modular but typically requires more complex training schemes and engineering overhead~\cite{zhang2021end}. The second takes an end-to-end strategy, trading modularity for simpler design and training schemes~\cite{mallidi2018device, howard2021neural, braun2022task, cornell2023implicit}.

\begin{figure}[!t]
    \centering
    \includegraphics[trim=5mm 4mm 5mm 2mm, clip, width=1\linewidth]{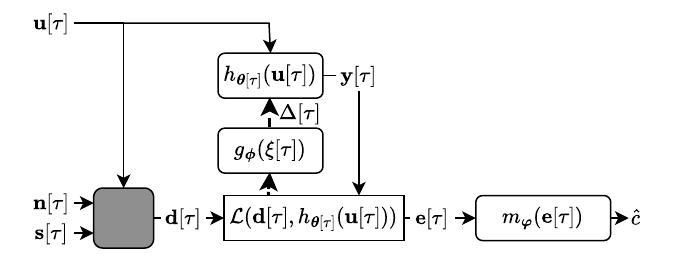}
    \caption{The echo canceller $h_{\btheta[\tau]}$, controlled by optimizer $g_{\bphi}$, cancels own-playback and passes its output to keyword classifier $m_{\bvarphi}$. The optimizer is trained end-to-end with the classifier, improving performance without the need for additional tuning.}
    \label{fig:block_diagram}
\end{figure}

In this paper, we propose a novel framework that bridges these two DNN-based approaches, allowing for the incorporation of downstream classification task knowledge into AFs without task-dependent design or test-time feedback. We propose a modified training scheme for the Meta-AF framework~\cite{casebeer2022meta}, which we call classification-trained Meta-AF~(CT-Meta-AF). CT-Meta-AF is unique in that it does not require oracle signal-level supervision such as impulse responses or clean speech. Instead, it leverages a combination of signal-level self-supervision and automatic classifier feedback. This eliminates the need for manual task/model dependent tuning and is directly compatible with both off-the-shelf and end-to-end trained downstream models.

To evaluate our framework, we apply it to two fundamental tasks for smart devices: acoustic echo cancellation~(AEC) and keyword spotting~(KWS). AEC is a critical component of nearly all devices with own-playback, and KWS is essential for enabling hands-free operation of smart devices. We construct a challenging joint AEC and KWS dataset by combining two existing datasets~\cite{warden2018speech, cutler2022AEC}, and benchmark CT-Meta-AFs against Meta-AF~\cite{wu2022meta} and a Kalman Filter~\cite{kuech2014state}, two state-of-the-art approaches for signal-level objectives. Results show that CT-Meta-AF achieves consistent performance gains across both synthetic and real playback scenarios, multiple playback lengths, various task/model configurations, and both pre-trained and joint-trained keyword spotting models.

The contributions of this work are:
1) A versatile framework for seamlessly integrating downstream task knowledge into AF update rules, applicable to various AFs and tasks,
2) Task-specific insights for AEC and KWS, yielding models that surpass previous approaches,
3) A comprehensive empirical exploration of incorporating downstream task knowledge across different classifier models, employing various training schemes, and evaluating on both real-world and synthetic data. We also release all code and weights$^1$.

%% file: sections/background.tex
\def\time{{\mathrm{t}}}
\def\freq{\mathrm{k}}
\def\mic{{\mathrm{m}}}
\def\buffer{\mathrm{b}}
\def\frame{\tau}

\def\F{\mathbf{F}}
\def\I{\mathbf{I}}
\def\0{\mathbf{0}}
\def\1{\mathbf{1}}

\def\u{\mathbf{u}}
\def\U{\mathbf{U}}
\def\d{\mathbf{d}}
\def\D{\mathbf{D}}
\def\y{\mathbf{y}}
\def\e{\mathbf{e}}
\def\s{\mathbf{s}}
\def\w{\mathbf{w}}

\def\H{\mathsf{H}}
\def\T{\top}
\def\diag{\operatorname{diag}}

\def\L{\mathcal{L}}
\def\grad{\bm{\nabla}}

\section{Meta-AF Background}

\subsection{Adaptive Filters}
\label{sec:background}
In this work, an AF is a time-varying filtering procedure with parameters that are controlled by an online optimization rule in order to solve some objective function. We denote the filter as $h_{\btheta[\frame]}$, parameterized by $\btheta[\frame]$, and updated using an additive rule,
\begin{equation}
    \btheta[\frame + 1] = \btheta[\frame] + \bm{\Delta}[\frame].
\label{eq:general_update}
\end{equation}
Time-varying updates $\bm{\Delta}[\frame]$ govern filter adaptation, and are produced by an update rule $g_{\bphi}$. Conventional AFs are designed to produce updates which solve a tractable AF loss $\L(\cdot)$, which is typically a function of the AF output and some readily available mixture.

To perform our chosen AF task of AEC, we fit a filter to to remove echo caused by own-playback. The time-domain signal model is $\underline{\d}[\time]=\underline{\u}[\time]\ast\underline{\w}+\underline{\mathbf{n}}[\time]+\underline{\mathbf{s}}[\time]$, where $\underline{\mathbf{n}}$ is noise, $\underline{\mathbf{s}}$ is a keyword, and $\underline{\w}$ is the transfer function applied to own playback $\underline{\mathbf{u}}$. The AF attempts to mimic $\underline{\w}$ via a multi-frame frequency-domain filter applied using overlap-save~\cite{soo1990multidelay}. We denote this as $h_{\btheta[\frame]}$ with frequency coefficients $\btheta[\frame] = \hat{\w}[\frame]\in \mathbb{C}^{K\times B}$ where $K$ is frequencies, $B$ is the number of blocks, and blocks advance by $K/2$ samples. The output is echo reduced signal $\underline{\e}[\frame] = h_{\btheta[\frame]}(\underline{\d}[\frame], \underline{\u}[\frame])$.

\subsection{Meta-Adaptive Filters}
The Meta-AF~\cite{casebeer2022meta} approach showed that AF control rules could be entirely learned with a learning-to-learn strategy~\cite{andrychowicz2016learning}. Instead of solving the AF loss $\L(\cdot)$ directly, the procedure solves a meta-loss,
\begin{equation}
    \hat{\bphi} = \arg\min_{\bphi} E_\mathcal{D}[\;  \L_M(\; g_{\bphi},  \L(h_{\btheta}, \cdots) \; ) \; ], 
\label{eq:meta}
\end{equation}
where $\L_M(\; g_{\bphi}, \L(h_{\btheta}, \cdots) \;)$ is the meta-loss that is a function of the AF $h_{\btheta[\frame]}$, optimizer neural network $g_{\bphi}$, and $E_\mathcal{D}$ represents expectation over dataset $\mathcal{D}$. The standard setup solves this via an additive update $\btheta[\frame + 1] = \btheta[\frame] + g_{\bphi}(\cdot)$,
where $g_{\bphi}$ is a complex valued DNN  parameterized by $\bphi$, applied per frequency $\freq$. Typically, $g_{\bphi}$ has per-frequency state and inputs,
\begin{eqnarray}
(\bm{\Delta}_{\freq}[\frame], \bm{\psi}_{\freq}[\frame+1]) &=& g_{\bphi}(\bm{\xi}_{\freq}[\frame], \bm{\psi}_{\freq}[\frame]) \label{eq:full_neural_update1}\\
\btheta_{\freq}[\frame + 1] &=& \btheta_{\freq}[\frame] +  \bm{\Delta}_{\freq}[\frame],\label{eq:full_neural_update2}
\end{eqnarray}
where we index across $K$ frequencies using subscript $\freq$. The input is a stack of the frequency-domain filter gradient, input, target, error, and output respectively: $\bm{\xi}_{\freq}[\frame] = [\nabla_{\freq}[\frame], \u_{\freq}[\frame], \d_{\freq}[\frame], \e_{\freq}[\frame], \y_{\freq}[\frame]]$. The outputs are AF update $\bm{\Delta}_{\freq}[\frame]$ and a new internal state $\bm{\psi}_{\freq}[\frame]$. To learn parameters $\bphi$, we use backpropagation-through-time and a meta-optimizer~(e.g. Adam) over $L$ optimization steps. We set our meta-loss to the self-supervised log-mse meta-loss,
\begin{eqnarray}
    \L_M(\underline{\bar{\e}}) &=& \ln E[\|\underline{\bar{\e}}[\frame]\|^2],
     \label{eq:metaloss}
\end{eqnarray}
where $\underline{\bar{\e}}[\frame] = \mathrm{cat}(\underline\e[\frame], \cdots, \underline\e[\frame+L-1]) \in \mathbb{R}^{RL}$, and $\mathrm{cat}$ is the concatenation operator. Intuitively, this loss encourages an optimizer to maximally reduce the energy of the AF output without needing oracle echo~($\underline\u \ast \underline\w$) or oracle speech~($\underline\s$) information.

%% file: sections/methods.tex
\section{Classification-Trained Meta-AF}
In this work, we propose CT-Meta-AF, an extension to the Meta-AF framework~\cite{casebeer2022meta}. CT-Meta-AF learns an AF optimizer for downstream classification tasks in a data-driven manner, without task/model dependent engineering or the need for oracle signals like clean speech. Specifically, we use an adaptive filter, such as an echo canceller, whose time-varying parameters $\btheta[\tau]$ are updated by a meta-learned update rule $g_{\bphi}$. The goal is to use the processed signals from the adaptive filter for a downstream classification task, such as keyword spotting, performed by a model $m_{\bvarphi}$. To this end, we employ an end-to-end training setup that enables joint optimization of the adaptive filter, optimizer, and the classification model. 
Our approach uses feedback from the classifier during training to learn a custom optimizer, resulting in enhanced performance without requiring explicit coupling or the need for test-time feedback.
This approach is also suitable for training on real mixtures. 

\subsection{Overview}
To address the downstream objective, we introduce a modification to the original meta-loss from \eqref{eq:metaloss} that incorporates classification feedback, which we call $\L_{CM}$. Specifically, $\L_{CM}$ consists of two components: the classification loss, denoted by $\mathcal{L}_{C}(c, \hat{c})$, where $c$ is the true class and $\hat{c}=m_{\bvarphi}(\cdot)$ is the predicted class, and the self-supervised loss, denoted by $\L_M$, as defined in \eqref{eq:metaloss}. The joint loss,
\begin{equation}
\L_{CM}(\bar{\e}, c, \hat{c}) = \lambda \cdot \L_C(c, \hat{c}) + (1 - \lambda) \cdot \L_M(\bar{\e}),
\label{eq:meta_cls_loss}
\end{equation}
uses $\lambda \in [0,1]$ to control the weighting between classification and self-supervised losses. This approach implicitly couples the representations learned by $g_{\bphi}$ and $m_{\bvarphi}$ without explicitly sharing their parameters, using the idea of ``task-splitting"~\cite{braun2022task, cornell2023implicit}.

\subsection{Pretrained Classifier}
\label{sec:pre_32}
When dealing with complex classification tasks, it is common to use off-the-shelf or pretrained models due to data, compute, or engineering constraints. In this setting, we assume that the classifier, $m_{\bvarphi}$, is frozen and not trainable. This leads to a modified \eref{eq:meta}, 
\begin{equation}
\hat{\bphi} = \arg\min_{\bphi} E_\mathcal{D}[\; \L_{CM}(\; m_{\bvarphi}, g_{\bphi}, \L(h_{\btheta}, \cdots) \; ) \; ].
\label{eq:meta_cls_pre}
\end{equation}
This equation incorporates $m_{\bvarphi}$ without modifying it by using $m_{\bvarphi}$ as an additional loss term. This method allows custom training of the optimizer for the classifier without the classifier needing to be aware of the custom preprocessing. It customizes the optimizer to the classifier, automatically performing the model unification proposed by Seltzer et al.~\cite{seltzer2004likelihood} in a model and task independent fashion.

\subsection{Jointly Trained Classifier}
\label{sec:joint_33}
When ample data and compute resources are available, joint or end-to-end training is a powerful approach. This involves training both the optimizer $g_{\bphi}$ and classifier $m_{\bvarphi}$ simultaneously using the modified meta-loss function $\L_{CM}$. The parameters of the optimizer and classifier, $\bphi$ and $\bvarphi$, are jointly optimized via 
\begin{equation}
\hat{\bphi}, \hat{\bvarphi} = \arg\min_{\bphi, \bvarphi} E_\mathcal{D}[\; \L_{CM}(\; m_{\bvarphi}, g_{\bphi}, \L(h_{\btheta}, \cdots) \; ) \; ].
\label{eq:meta_cls_joint}
\end{equation}
During joint training, classification feedback is used to adjust the parameters of the optimizer, improving its ability to generate outputs that aid in classification. At the same time, the classifier learns to operate on AF outputs. This builds upon the pretrained classifier setup by producing a custom-trained classifier that can better leverage the AF. To speed up training, we can initialize $\bphi$ and $\bvarphi$ with pretrained weights. It is also possible to improve performance by allowing $m_{\bvarphi}$ to use information beyond the outputs of $h_{\btheta}$.

%% file: sections/experimental_setup.tex
\section{Experimental Design}
The goal of our setup is to demonstrate that CT-Meta-AF strikes the right balance between flexibility and structure, enabling high-performance while still being a direct replacement for existing approaches. We evaluate the performance of CT-Meta-AF on KWS with device playback. The AEC removes own-playback from recordings by learning the echo path, as shown in \fref{fig:block_diagram}. For model details, see \fref{fig:aec_kws_detail}. We evaluate three scenarios: one with a pretrained KWS, one with test-time swapped KWS, and one with a jointly trained KWS, all implemented using Meta-AF~\cite{casebeer2022meta}.
All CT-Meta-AFs use the classifier for training. The jointly trained KWS is optimized end-to-end and trained simultaneously with the AEC.

\subsection{AEC Configurations}
To perform AEC, we fit an AF to cancel echo. At inference time, we only assume access to the playback, $\underline{\u}$ and echoic mixture, $\underline{\d}$. The AF uses frequency-domain overlap save with a window of $K=1024$, a hop of $R=512$, and $B=4$ blocks. Each block is denoted by $\mathbb{C}\text{Filter}$ in \fref{fig:aec_kws_detail}. Each AEC filters the farend $\u[\frame]$ with estimated filter $\hat\w[\frame]$ and subtracts the result from $\d[\frame]$. For a review of AEC see Benesty et al.~\cite{benesty2001advances}. We compare our approach to four baselines: regular Meta-AF~\cite{casebeer2022meta}, a Kalman filter AEC~(Diag. KF)~\cite{kuech2014state}, No-Echo, and No-AEC. We grid-search-tune Diag. KF using KWS accuracy. No-Echo simulates running the KWS model in a playback-free environment, while No-AEC performs no echo cancellation and simulates deploying a KWS without AEC.

CT-Meta-AF and Meta-AF use the same higher-order architecture~\cite{wu2022meta} and training scheme, with the only difference being the loss. For CT-Meta-AF, we use $\lambda=0.5$ in \eref{eq:meta_cls_loss}, and for regular Meta-AF, we use $0$. We set the group size, group hop, and hidden size to $5$, $2$, and $48$, respectively~(see \fref{fig:aec_kws_detail} for details). We use Adam with a batch size of $16$, a learning rate of $2\cdot 10^{-4}$, momentum $\beta_1=0.99$, and we randomize the truncation length $L$. Additionally, we apply gradient clipping and reduce the learning rate by half if the validation performance does not improve for $10$ epochs, and we stop training after $30$ epochs with no improvement. For pretrained KWS experiments, we train $\bphi$ from scratch, and for joint training, we initialize with the best pretrained $\bphi$. When joint training with Diag. KF or Meta-AF we do not update the pretrained $\bphi$ and just train the KWS. Each model has $32\,$K complex parameters and trains on one GPU.
\subsection{KWS Configurations}
\begin{figure}[!t]
    \centering
    \includegraphics[trim=6mm 1mm 9mm 12mm, clip, width=1\linewidth]{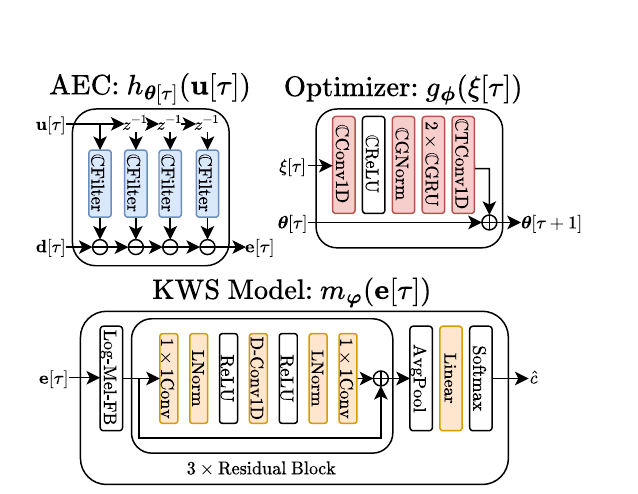}
    \caption{Detailed view of the AEC, Optimizer, and KWS. The AEC preprocesses the KWS inputs and the optimizer controls the AEC parameters. At test time, only the AEC parameters change}
    \label{fig:aec_kws_detail}
\end{figure}

We set up KWS as a multi-class classification task, where the keyword $\underline{\mathbf{s}}$ belongs to a single class $c$. We base our KWS on the model proposed by Cornell et al.~\cite{cornell2023implicit}. Our version uses $40$ log-mel-filter-bank inputs and short-time Fourier transform with a $256$ point hop and a $512$ point window. The KWS has $3$ residual blocks, each with a $1\times1$ convolution, layer norm, ReLU, a dilated convolution with kernel size $5$, layer norm, ReLU, and a final $1\times1$ convolution. The last residual block is averaged across time and fed to a dense layer with softmax to predict the class distribution. See \fref{fig:aec_kws_detail}.  We use a binary cross-entropy loss and train the KWS for $50$ epochs with a batch size of $128$, a learning rate of $10^{-3}$, and use the best checkpoint based on validation performance. We pretrain on a dataset without playback. For joint training, due to limited GPU capacity, we use a batchsize of $16$, learning rate of $10^{-4}$, $\beta_1=.9$, reduce the learning rate by half after $10$ epochs with no improvement, and stop training after $50$ epochs with no improvement. All KWS models have $\approx300$K parameters and use $\approx20$MFLOPs per second.

\subsection{Dataset \& Metrics}
We use synthetic playback during training and evaluate our models using both synthetic and real playback data from the Microsoft AEC Challenge~\cite{cutler2022AEC} and keywords from the Google Speech Commands V2 dataset~\cite{warden2018speech}, sampled at $16\,$KHz. During training, we trim playback to $3$ seconds, randomly mix playback with keywords, zero-pad keywords as needed, apply a random shift, and set the signal-to-echo ratio~(SER) uniformly at random between $-25\,$dB and $0\,$dB. Each keyword is used once. To test, we use playback trimmed to either $4$ or $12$ seconds. We evaluate three datasets: a $35$-class dataset and two smaller datasets with $10$ and $2$ classes, respectively. The smaller datasets contain approximately one-third and one-tenth of the full dataset. We always use the same folds and never mix across folds.

To evaluate performance, we use macro and micro averaged F1, where higher is better. The goal is to reflect the class imbalance in the KWS Dataset. We display results as macro F1~(micro F1).

%% file: sections/results.tex
\section{Results \& Discussion}
Here, we investigate the effect of classification training for Meta-AF on AEC with a downstream KWS. This task simulates a device attempting to classify a spoken keyword while emitting playback. We show results for zero playback~(No-Echo), no cancellation~(No-AEC), Diag. KF, Meta-AF based AEC without classification training~(Meta-AEC), and Meta-AF based AEC with classification training~(CT-Meta-AEC). The Diag. KF and Meta-AEC baselines are state-of-the-art traditional and data-driven approaches. All models are trained on synthetic playback with challenging SERs distributed uniformly at random between $-25\,$dB and $0\,$dB, and real playback is only used for evaluation. We use an off-the-shelf KWS that is not customized for AEC in sections \ref{sec:pre} and \ref{sec:mod} but do experiment with retraining the KWS in section \ref{sec:joint}. We evaluate longer playback and compute statistical significance for select models. Across experiments, CT-Meta-AF proved highly stable and required no additional tuning, making it a drop-in replacement for prior approaches.

\subsection{Pretrained KWS}
\label{sec:pre}

First, we test the effectiveness of classification training as described in \ref{sec:pre_32}. In \tref{tab:aec_pre}, all AEC models share a frozen KWS model pretrained on keywords without playback. This setup mimics using an off-the-shelf KWS without retraining but with access to your own dataset, a common real-world setup. For CT-Meta-AEC training, we use $\lambda = 0.5$ when computing $\L_{CM}$. The F1 degradation from No-Echo to No-AEC shows the importance of AEC.

In \tref{tab:aec_pre}, CT-Meta-AEC significantly outperforms its signal-level Diag. KF and Meta-AEC counterparts, as indicated by better F1 scores. In synthetic playback~(left column), CT-Meta-AEC improves over Meta-AEC by some .165 F1. This gap persists in real playback, where CT-Meta-AEC is the top performing model by .091 F1. Interestingly, CT-Meta-AEC achieves a 5.80 dB reduction in echo, while regular Meta-AF attains a 9.57 dB reduction. This observation underscores the limitations of signal-level metrics as effective descriptors of downstream performance, especially in the context of nonlinear processing methods such as DNN-based KWS. Classification-based training effectively addresses this discrepancy and optimizes downstream performance. In a secondary evaluation with 12-second signals, CT-Meta-AEC outperforms Meta-AEC by 0.279 F1 in synthetic playback and 0.119 F1 in real playback.

\subsection{Specialization and Sensitivity}
\label{sec:mod}

Next, we study the specialization capabilities of CT-Meta-AEC in different KWS setups. We use the approach from \ref{sec:pre_32} to train three CT-Meta-AEC models: CT-Meta-AEC-35, CT-Meta-AEC-10, and CT-Meta-AEC-2, corresponding to datasets with 35, 10, and 2 keyword classes, respectively. We assess the performance of each CT-Meta-AEC model on all three datasets, but always use a downstream KWS model specifically trained on a matching keyword setup. \tref{tab:aec_swap} presents the results of our experiments on synthetic playback scenarios, where each row corresponds to a distinct AEC model and each column represents a unique test-time KWS configuration. The first row displays our Meta-AEC baseline. Table entries on the diagonal show matched train/test setups.

In all columns of \tref{tab:aec_swap}, CT-Meta-AEC with a matching KWS performs best. This demonstrates that CT-Meta-AEC is learning a specialized optimizer~(the diagonal) which outperforms both generalist optimizer~ (top row) and optimizers trained on different models~(off diagonal). Notably, results above the diagonal show that specialization can yield superior results even in the presence of limited training data. Specifically, Meta-AEC and CT-Meta-AEC-35, which are trained with three times more data than CT-Meta-AEC-10, are surpassed by the latter on 10 class KWS. This suggests that classification training is a data-efficient approach for improving performance on downstream tasks. The results from mismatched train/test setups show that CT-Meta-AECs are good at related tasks, and usually still outperform Meta-AEC.

\subsection{Jointly Trained KWS}
\label{sec:joint}
We investigate joint training of the AEC and KWS to simulate having sufficient data for training a complete system, without oracle signal-level supervision. Both the AEC and KWS models undergo joint training, following the approach in \ref{sec:joint_33}.The KWS models undergo a pre-training phase on scenes devoid of playback, followed by fine-tuning using the output of their corresponding AEC models. This represents a significant departure from the approach outlined in \ref{sec:pre_32}, where KWS models were trained without consideration for echo presence. We call this approach JCT-Meta-AEC. For statistical robustness, we ran three separate trials of the Meta models. 

In \tref{tab:aec_joint}, we find that the training scheme from \ref{sec:joint_33} improves performance. JCT-Meta-AEC outperforms Meta-AEC by $.008$ F1~(p-value $.01$) in real playback scenarios, underscoring the efficacy of classification training for real-world performance. While JCT-Meta-AEC outperforms Meta-AEC on synthetic data, the improvement is not significant. However, on the longer $12$-second test-set, JCT-Meta-AEC beats Meta-AEC by a margin of 0.057~(p-value $.0005$). Of note, we did not encounter any training stability issues.

\begin{table}[!tb]
    \centering
    \begin{tabular*}{.99\linewidth}{@{\extracolsep{\fill}}c c c c c@{}}\toprule
        \textit{Model} && \textit{Synthetic Playback} && \textit{Real Playback}\\ \hline
        No-Echo && .928~(.935) && .931~(.936)\\
        No-AEC && .087~(.098) && .102~(.117)\\
        Diag. KF && .196~(.208) && .165~(.180)\\
        Meta-AEC && .335~(.340) && .226~(.236)\\
        CT-Meta-AEC && \textbf{.500}~(\textbf{.508}) && \textbf{.317}~(\textbf{.326}) \\
        \bottomrule
    \end{tabular*}
    \caption{F1 Macro~(F1 Micro) with a pretrained KWS and SER $\in[-25, 0]\,$dB. No-AEC and No-Echo are lower/upper bounds.}
    \label{tab:aec_pre}
\end{table}

\begin{table}[!tb]
    \centering
    \begin{tabular*}{.99\linewidth}{@{\extracolsep{\fill}}c c c c@{}}\toprule
        \textit{Model} & \textit{35 Class} & \textit{10 Class} & \textit{2 Class}\\\hline
        Meta-AEC & .335~(.340) & .476~(.465)  & .751~(.754)\\
        CT-Meta-AEC-35 & \textbf{.500}~(\textbf{.508})  & .513~(.483) & .786~(.788)\\
        CT-Meta-AEC-10 & .344~(.348)  & \textbf{.533}~(\textbf{.529}) & .781~(.781) \\
        CT-Meta-AEC-2 & .274~(.281) & .413~(.400)  & \textbf{.802}~(\textbf{.802})\\
        \bottomrule
    \end{tabular*}
    \caption{Specialization of classification-trained Meta-AEC with performance shown as F1 Macro~(F1 Micro). A different model with different keywords is swapped in at test time. SER $\in[-25, 0]\,$ dB.}
    \label{tab:aec_swap}
\end{table}

\begin{table}[!tb]
    \centering
    \begin{tabular*}{.99\linewidth}{@{\extracolsep{\fill}}c c c c c@{}}\toprule
        \textit{Model} && \textit{Synthetic Playback} && \textit{Real Playback}\\ \hline
        No-Echo && .922~(.928) && .931~(.936)\\
        No-AEC && .609~(.619) && .583~(.593)\\
        Diag. KF && .778~(.781) && .690~(.698)\\
        Meta-AEC* &&.870~(.876) && .746~(.754)\\
        JCT-Meta-AEC* && \textbf{.871}~(\textbf{.877}) && \textbf{.754}~(\textbf{.762})\\
        \bottomrule
    \end{tabular*}
    \caption{F1 Macro~(F1 Micro) with a jointly trained KWS and SER $\in[-25, 0]\,$dB. No-AEC and No-Echo are lower/upper bounds. An asterix denotes that we averaged three trials.}
    \label{tab:aec_joint}
\end{table}

%% file: sections/conclusion.tex
\section{Conclusion}
We proposed a simple yet effective approach for improving the performance of downstream classification tasks by incorporating classifier feedback into learned adaptive filter update rules. Our approach, classification-trained meta-adaptive filtering~(CT-Meta-AF) enables end-to-end training without oracle single-level supervision. We evaluated CT-Meta-AF on echo cancellation and keyword spotting and observed consistent performance gains across synthetic and real playback, multiple keyword configurations, and both pre and jointly trained keyword models. CT-Meta-AF yields performance improvements without additional inference costs or manual tuning. We believe our approach has the potential to improve many adaptive filter pipelines, thanks to it's plug-and-play design and promising real-world performance.